# Epimenides, Gödel, Turing:
# an Eternal Gölden Tangle [0]

Eric C.R. Hehner

Department of Computer Science, University of Toronto
hehner@cs.utoronto.ca

**Abstract:** The Halting Problem is a version of the Liar's Paradox.

## Epimenides

An ancient Cretan named Epimenides is reported to have said "All Cretans are liars." [1]. This is supposed to be self-contradictory, but it misses the mark. If there is any other Cretan, and that Cretan is a truth-teller, then Epimenides' sentence is simply false: Epimenides is a liar, but not all Cretans are liars. St. Paul missed the point completely, taking Epimenides' statement at face value, and elaborating: "It was one of themselves, one of their own prophets, who said, "Cretans were never anything but liars, dangerous animals, and lazy":  and that is a true statement." [5]. I will refer to the simpler sentence
> This sentence is false.

as the Liar's Paradox. If that sentence is true, then, according to the sentence, it is false. If it is false, then it is true. That simple sentence is self-contradictory.

I give the sentence a name, say $L$ for Liar.
> $L$:  $L$ is false.

As a mathematical formula, it becomes
> $L = (L\textit{=false})$

As an equation in unknown $L$, it has no solution, because the equation is *false* regardless of whether $L$ is *true* or *false*. As a definition or specification of $L$ it is called "inconsistent". (I am using italic *true* and *false* for the boolean constants representing truth and falsity.)

A slightly more complicated version presents the inconsistency as two sentences.
> The next sentence is true.
> The previous sentence is false.

Naming the first sentence $B$ and the second $G$, as mathematical formulas, they become
> $B = (G\textit{=true})$
> $G = (B\textit{=false})$

These two equations in the two unknowns $B$ and $G$ have no solution: there is no assignment of boolean values to $B$ and $G$ that satisfies the two equations. They are inconsistent. If you look at either one of the sentences alone, there is no inconsistency. It may make sense to say that the next sentence is true, and it may make sense to say that the previous sentence is false. But together they are inconsistent.

Let me complicate this inconsistency by adding a parameter, so $B$ can say whether any sentence is true, not just sentence $G$. To reduce contention over truth and falsity, I will stick with mathematical sentences, otherwise known as boolean expressions (allowing subexpressions of any type, including functions). To pass sentences as data, we need to encode them in some way. The easiest encoding is as a character string. Now $B$ becomes a function from strings to booleans, and the pair of sentences become
> $B\,(s)\ =\ \textit{true}$ if string $s$ represents a boolean expression with value *true* ;
>     *false* otherwise
> $G\ =\ $" $B\,(G) = \textit{false}$ "

I have made two definitions: *B* and *G* . Since *G* is just a character string, there cannot be anything wrong with its definition; it represents the boolean expression *B* (*G*) = *false* . But the definition of *B* , no matter how carefully worded, no matter how clear it sounds, conceals an inconsistency. I am not concerned with computing *B* ; I just want to define a mathematical function. The parameter allows us to show a large number of examples, like *B* ("0=0") = *true* and *B* ("0=1") = *false* , which are not problematic. They may fool us into believing that the definition of *B* makes sense. But they are irrelevant. The inconsistency is revealed by applying *B* to *G* . If *B* (*G*) = *true* , then *G* represents a *false* expression, so *B* (*G*) should be *false* . If *B* (*G*) = *false* , then *G* represents a *true* expression, so *B* (*G*) should be *true* . The inconsistency is the same as in the unparameterized, unencoded version of the Liar's Paradox.

## Gödel

The Liar's Paradox is about truth. Gödel used the same self-contradictory construction to talk about provability [2]. He used a numeric, rather than string, encoding of sentences, and he used the name *Bew* (short for Beweisbar, which is German for provable) for a function similar to *B* . The sentence encoded by *G* is popularly called "the Gödel sentence". With our notations and encoding, *B* and *G* become

>  *B* (*s*)  =  *true*  if string  *s*  represents a provable boolean expression;
>             *false*  otherwise
>  *G*  =  " *B* (*G*) = *false* "

Now we ask: Is *B* (*G*) = *true* ? If we suppose *B* (*G*) = *true* , then *G* represents a *false* sentence, and in a consistent logic, no *false* sentence is provable, so *B* (*G*) should be *false* . If we suppose *B* (*G*) = *false* , then *G* represents a *true* sentence, and in a complete logic, all *true* sentences are provable, so *B* (*G*) should be *true* . Gödel concluded that if a logic is expressive enough to define *B* , then the logic is either inconsistent or incomplete.

## Turing

Epimenides talked about truth; Gödel talked about provability; Turing talked about computability using the same sort of arguments [4]. For my examples, I will use the Pascal programming language, but the choice of language is irrelevant; any other general-purpose programming language would do just as well. I'll start with a procedure named *liar* that is closely analogous to the Liar's Paradox.

**procedure** *liar*;
**begin**
    **if** (execution of *liar* terminates) **then** *liar*
**end**

I have not finished writing procedure *liar* ; what remains is to replace the informal boolean expression (execution of *liar* terminates) with either *true* or *false* , whichever one is appropriate. The problem in doing so is that the informal boolean expression refers to itself in a self-contradictory manner: if the execution of procedure *liar* terminates, it should be replaced with *true* , creating a procedure whose execution does not terminate; if the execution of *liar* does not terminate, it should be replaced with *false* , creating a procedure whose execution does terminate. This is not a programming problem, not a computability problem, not a lack of expressiveness of Pascal. The problem is that the informal boolean expression is an inconsistent specification. One might protest:
> Either execution of *liar* terminates, or it doesn't. If it terminates, use *true* ; if it doesn't, use *false* . How can there possibly be an inconsistency?

But I hope the inconsistency is clear enough that no-one will protest.

As we did with the Liar's Paradox, let's present the same inconsistency as two declarations.

**function** *halts*: **boolean**;
**begin**
    { return *true* if execution of *diag* terminates; }
    { return *false* otherwise }
**end**;

**procedure** *diag*;
**begin**
    **if** *halts* **then** *diag*
**end**

Only the header of function *halts* appears. In place of the body there is a comment to specify what the body should be. If execution of procedure *diag* terminates, then the body should be *halts*:= *true* . If execution of procedure *diag* does not terminate, then the body should be *halts*:= *false* . So there is no problem in programming the body. The problem is to decide whether execution of *diag* terminates. If we suppose it does, then *halts* should return *true* , and so we see that execution of *diag* does not terminate. If we suppose it does not, then *halts* should return *false* , and so we see that execution of *diag* does terminate.

Procedure *diag* has been written in its entirety. Syntactically, it is a procedure; to determine that *halts* is being used correctly within *diag* , we need only the header for *halts* , not the body, and we have the header. Semantically, it is a procedure; to determine the meaning of the call to *halts* within *diag* , we need only the specification of *halts* , not its implementation, and we have the specification. (That important programming principle enables a programmer to call procedures written by other people, knowing only the specification, not the implementation. It also enables a programmer to change the implementation of a procedure, but still satisfying the specification, without knowing where and why the procedure is being called.) So there is nothing wrong with the definition of *diag* . The problem is that we cannot write the body of *halts* to satisfy its specification. This is not a programming problem, not a computability problem, not a lack of expressiveness of Pascal. The problem is that the specification of the body of *halts* is inconsistent. One might protest:
    Either execution of *diag* terminates, or it doesn't. If it terminates, use *halts*:= *true* ; if it doesn't, use *halts*:= *false* . How can there possibly be an inconsistency?
The inconsistency cannot be seen by looking only at *halts* or only at *diag* . Each refers to the other, and together they are inconsistent.

Let me complicate this inconsistency by adding a parameter, so *halts* can say whether execution of any parameterless Pascal procedure terminates, not just *diag* . To pass procedures as data, we need to encode them in some way, and the easiest encoding is as a character string. (Whenever programs are presented as input data to a compiler or interpreter, they are presented as character strings.) We assume there is a dictionary of function and procedure definitions that is accessible to *halts* , so that the call *halts* ('*diag*') allows *halts* to look up '*diag*' , and subsequently '*halts*' , in the dictionary, and retrieve their texts for analysis.

**function** *halts* (*p*: **string**): **boolean**;
{ return *true* if *p* represents a parameterless Pascal procedure whose execution terminates; }
{ return *false* otherwise }

**procedure** *diag*;
**begin**
    **if** *halts* ('*diag*') **then** *diag*
**end**

To determine that *diag* is syntactically a Pascal procedure, we need only the header for *halts* , not the body, and we have the header. To determine the semantics of *diag* , we need only the specification of *halts* , not its implementation, and we have the specification.

As before, we cannot write the body of *halts* to satisfy the specification. No matter how carefully worded it is, no matter how clear it sounds, the specification conceals an inconsistency. The inconsistency is revealed by applying *halts* to '*diag*' . If *halts* ('*diag*') = *true* , then execution of *diag* is nonterminating, so *halts* ('*diag*') should be *false* . If *halts* ('*diag*') = *false* , then execution of *diag* is terminating, so *halts* ('*diag*') should be *true* . This is still not a programming problem, not a computability problem, not a lack of expressiveness of Pascal. It is still the same inconsistency that was present in the unparameterized, unencoded version, and the same inconsistency that was present in the *liar* procedure. One might protest:
> Either execution of a procedure represented by *p* terminates, or it doesn't. If it terminates, *halts* (*p*) should return *true* ; if it doesn't, *halts* (*p*) should return *false* . How can there possibly be an inconsistency?

Now the protest starts to sound more plausible because the parameter allows us to show a large number of examples which are not problematic. For example,
> *halts* ('*stop*') = *true*
> *halts* ('*go*') = *false*

where *stop* and *go* are defined as
> **procedure** *stop*; **begin end**
> **procedure** *go*; **begin** *go* **end**

These nonproblematic examples may fool us into believing that the specification of *halts* makes sense. But they are irrelevant. Procedure *diag* shows us the inconsistency.

There is one last complication: a second parameter so *halts* can say whether execution of any Pascal procedure with an input parameter terminates.

**function** *halts* (*p*, *i*: **string**): **boolean**;
{ return *true* if *p* represents a Pascal procedure with one string input parameter }
{ whose execution terminates when given input *i* ; return *false* otherwise }

**procedure** *diag* (*s*: **string**);
**begin**
    **if** *halts* (*s*, *s*) **then** *diag* (*s*)
**end**

This is now a modern version of Turing's Halting Problem. Turing's argument is as follows.

> Assume that *halts* is computable, and that it has been programmed according to its specification. Does execution of *diag* ('*diag*') terminate? If it terminates, then *halts* ('*diag*', '*diag*') returns *true* , and so we see from the body of *diag* that execution of *diag* ('*diag*') does not terminate. If it does not terminate, then *halts* ('*diag*', '*diag*') returns *false* , and so from the body of *diag* , execution of *diag* ('*diag*') terminates. This is inconsistent. Therefore function *halts* cannot have been programmed according to its specification; *halts* is incomputable.

The two parameters (*p*, *i*) make a two-dimensional space, and point ('*diag*', '*diag*') is on its diagonal, which is why the argument is called a diagonal argument, and why I named the procedure *diag* . But any string would do equally well as a value for the second parameter, and the second parameter adds nothing to Turing's argument.

The surprise, and the main point of this paper, is that the computability assumption is unnecessary to the argument. Without assuming that *halts* is computable, I ask what the result of *halts* ('*diag*', '*diag*') should be. This is a question about the specification of *halts*. If it should be *true*, then the semantics of *diag* ('*diag*') is nontermination, so *halts* ('*diag*', '*diag*') should be *false*. If it should be *false*, then the semantics of *diag* ('*diag*') is termination, so it should be *true*. This is inconsistent. Therefore *halts* cannot be programmed according to its specification. But the problem is not incomputability; it is inconsistency of specification. It is the same inconsistency that was present in all previous versions, before I added the complications of parameters and encodings. It is just the Liar's Paradox in fancy clothing. In fact, Turing's argument could have been applied to procedure *liar* with equal (in)validity.

**procedure** *liar*;
**begin**
    **if** (execution of *liar* terminates) **then** *liar*
**end**

Assume that the expression (execution of *liar* terminates) is computable, and that it has been programmed according to its specification. Does execution of *liar* terminate? If it terminates, then (execution of *liar* terminates) is *true*, and so we see from the body of *liar* that its execution does not terminate. If it does not terminate, then (execution of *liar* terminates) is *false*, and so from the body of *liar* its execution terminates. This is inconsistent. Therefore the expression (execution of *liar* terminates) cannot have been programmed according to its specification; it is incomputable.

Turing's argument can be applied to any property of program execution. For example,

**procedure** *liar1*;
**begin**
    **if** (execution of *liar1* prints 'A' ) **then** *print* ('B') **else** *print* ('A')
**end**

Termination of execution of *liar1* is not in question: when (execution of *liar1* prints 'A' ) is replaced with either *true* or *false*, whichever is appropriate, execution of *liar1* terminates. The question is whether 'A' or 'B' is printed. Turing's argument says that the property "prints 'A' " is incomputable, and so is every property of program execution (except for the trivial "always *true*" and "always *false*" properties) [3]. But the problem is not incomputability; the problem is inconsistency of specification.

## Underdetermination

The Liar's Paradox, the Gödel sentence, and Halting Problem are all examples of inconsistency, which is also known as overdetermination. Here, "determination" means ruling out possible solutions: if we rule out all possible solutions, we have overdetermination; if we are left with more than one possible solution, we have underdetermination. An example is the sentence
    This sentence is true.
Whereas the Liar's Paradox can be neither true nor false, the sentence just written can be either true or false. Giving the sentence the name $U$ for underdetermined, it becomes the formula
    $U = (U=true)$
As an equation in unknown $U$, it has two solutions: both *true* and *false*.

Here is another example.
    $B(s)$ = *true* if string $s$ represents a provable boolean expression;
        *false* otherwise
    $H$ = " $B(H) = true$ "

Now we ask: Is $B(H) = true$ ? If we suppose $B(H) = true$ , then $H$ represents a *true* sentence, and in a complete logic, all *true* sentences are provable, so $B(H)$ should be *true* , as supposed. If we suppose $B(H) = false$ , then $H$ represents a *false* sentence, and in a consistent logic, no *false* sentence is provable, so $B(H)$ should be *false* , as supposed. The specification of $B$ is both overdetermined (for $G$ ) and underdetermined (for $H$ ).

Here is the final example.

**function** *halts* (*p*, *i*: **string**): **boolean**;
{ return *true* if *p* represents a Pascal procedure with one string input parameter }
{ whose execution terminates when given input *i* ; return *false* otherwise }

**procedure** *what* (*s*: **string**);
**begin**
        **if not** *halts* (*s*, *s*) **then** *what* (*s*)
**end**

That is the same *halts* specification as before; it says that the *halts* function will tell us whether the execution of a procedure terminates. What does it say about *what* ? If we suppose that *halts* ('*what*', '*what*') = *true* , we see from the body of *what* that its execution terminates, so that was the right supposition. If we suppose that *halts* ('*what*', '*what*') = *false* , we see from the body of *what* that its execution does not terminate, so again that was the right supposition. We have come to another inadequacy of the *halts* specification. The specification sounds just right: neither overdetermined nor underdetermined. But we are forced by the examples to admit that the specification is not as it sounds. In at least one instance ( *diag* ), the *halts* specification is overdetermined, and in at least one instance ( *what* ), the *halts* specification is underdetermined.

## Conclusion

Epimenides' conclusion was:
    It is inconsistent to ask for a function whose result is *true* for all and only those strings representing true sentences in a sufficiently expressive language.
Gödel's conclusion was:
    It is inconsistent to ask for a function whose result is *true* for all and only those strings representing provable sentences in a sufficiently expressive language.
Turing's conclusion should have been:
    It is inconsistent to ask for a function, written in a programming language, whose result is *true* for all and only those strings representing procedures, written in that same language, whose execution terminates.

If "incomputable" meant having an inconsistent specification, then *halts* would be incomputable. But "incomputable" does not mean "inconsistent". It means that a well-defined mathematical function, one with a consistent specification, cannot be computed using a Turing-Machine-equivalent programming language. That question has not been addressed.

## Conjecture

I conjecture that every consistent first-order specification is satisfied by a computable function. This conjecture is a bit like the Löwenheim-Skolem theorem that every consistent first-order theory has a countable model.

[other papers on halting](other papers on halting)